\documentclass[letter]{aa} 
\usepackage{graphicx}
\usepackage{txfonts}
\begin{document}
   \title{The H$\alpha$ line forming region of AB Aur spatially
   resolved at sub-AU with the VEGA/CHARA spectro-interferometer}


\author{K. Perraut\inst{1}, M. Benisty\inst{2}, D. Mourard\inst{3},
S. Rajabi\inst{2,4}, F. Bacciotti\inst{2}, Ph. B\'{e}rio\inst{3},
D. Bonneau\inst{3}, O. Chesneau\inst{3}, J.M.~Clausse\inst{3},
O. Delaa\inst{3}, A. Marcotto\inst{3}, A. Roussel\inst{3},
A. Spang\inst{3}, Ph. Stee\inst{3}, I. Tallon-Bosc\inst{5},
H. McAlister\inst{6,7}, T.~ten~Brummelaar\inst{6},
J. Sturmann\inst{6}, L. Sturmann\inst{6}, N. Turner\inst{6},
C. Farrington\inst{6} and P.J. Goldfinger\inst{6}}

   \institute{Laboratoire d'Astrophysique de Grenoble (LAOG),
         Universit\'e Joseph-Fourier, UMR 5571 CNRS, BP 53, 38041
         Grenoble Cedex 09, France \and INAF-Osservatorio Astrofisico di Arcetri, Largo E. Fermi 5,
         50125 Firenze, Italy \and Laboratoire Fizeau, OCA/UNS/CNRS
         UMR6525, Parc Valrose, 06108 Nice cedex 2, France \and Centro de Astrofisica,
         Universidade do Porto, 4150-752 Porto, Portugal
          \and
         Universit\'{e} de Lyon, Lyon, F-69003, France; Universit\'{e}
         Lyon~1, Observatoire de Lyon, 9 avenue Charles Andr\'{e},
         Saint Genis Laval, F-69230; CNRS, UMR 5574, Centre de
         Recherche Astrophysique de Lyon; Ecole Normale
         Sup\'{e}rieure, Lyon, F-69007, France \and Georgia State University, P.O. Box 3969, Atlanta GA
         30302-3969, USA \and CHARA Array, Mount Wilson Observatory,
         91023 Mount Wilson CA, USA }
         \offprints{Karine.Perraut@obs.ujf-grenoble.fr} \date{Received
         ...; accepted ...}

  \abstract
   {A crucial issue in star formation is to
   understand the  physical mechanism by which mass is  accreted onto
   and  ejected  by a  young  star. To
   derive key  constraints on the launching point of the jets and on
   the geometry of the winds, the visible spectro-polarimeter VEGA
   installed on the CHARA optical array can be an efficient means of probing the
   structure and the kinematics of the hot circumstellar gas at sub-AU.}
   {For the first time, we observed the Herbig~Ae star AB Aur
   in  the  H$\alpha$  emission  line, using the  VEGA  low  spectral
   resolution (R=1700) on two baselines of the array.}
   {
     We computed and calibrated the spectral visibilities of AB~Aur between 610~nm and
     700~nm in spectral bands of 20.4~nm. To simultaneously reproduce the line profile and the
   inferred  visibility around H$\alpha$, we used a 1-D radiative transfer code (RAMIDUS/PROFILER) that calculates level
   populations for hydrogen atoms in a spherical geometry and that
   produces synthetic spectro-interferometric observables.}
   {We clearly resolved AB Aur in the H$\alpha$ line and in a part of the continuum,
   even at the smallest baseline of 34~m. The small P-Cygni absorption feature is indicative of an outflow but
   could not be explained by a spherical stellar wind model. Instead, it favors a magneto-centrifugal X-disk
   or disk-wind geometry. The fit of the spectral visibilities from 610 to 700~nm
   could not be accounted for by a wind alone, so another component inducing a visibility
   modulation around H$\alpha$  needed  to  be
   considered. We thus considered a brightness asymmetry possibly caused by large-scale nebulosity or by the known spiral structures. }
   {Thanks to the unique capabilities of VEGA, we managed to simultaneously record for the first time
a spectrum at a resolution of 1700 and spectral visibilities in the visible range on a target as faint as
$m_{V}$ = 7.1. It was possible to rule out a spherical geometry for the wind of AB~Aur and provide realistic solutions to account for the H$\alpha$ emission compatible with magneto-centrifugal
acceleration. It was difficult, however, to determine the exact morphology of the wind because of the surrounding asymmetric ne\-bulosity. The study illustrates the advantages of optical
interferometry and motivates observations of other bright young stars in the same way
to shed light on the accretion/ejection processes.}


    \authorrunning {K. Perraut et al.}
   \titlerunning{Resolving the H$\alpha$ line forming region of AB Aur}
   \maketitle
\section{Introduction}
The class of Herbig Ae/Be (HAEBE) stars gathers the pre-main-sequence stars
of intermediate  mass (1.5 $\leq  M/M_{\sun} \leq$ 10),  with spectral
types of B to F8, strong emission lines, and the presence of infrared to
submillimeter excess flux. Like  their solar-type analogs, the T Tauri
objects, HAeBe  objects are known to be surrounded  by protoplanetary
disks of gas and dust, responsible for this excess emission.
In the case of lower mass stars, the formation is commonly explained by
the gravitational collapse of a dusty disk cloud and magnetically controlled
accretion via a disk. The formation of more massive stars is still very uncertain.
However, determining if similar processes are at work
in these objects is extremely important, as it would help to esta\-blish
whether the accretion/ejection mechanism is universal irrespective of the mass of
the central body. 
The hot gas that can be involved in accretion
and  ejection flows  close to  the source and can be  used to  probe the
corresponding  physical  conditions  through  its emission  in  lines, especially those
in the optical range that have a high diagnostic potential. These phenomena,
however, occur in a small region of a few astronomical units (AU) around the star,
corresponding,   for   close  star   formation   regions,   to  a   few
milliseconds of arc. Such  a small scale can only be resolved with
interferometry.

Guided by these considerations, we have attempted for the first time
to observe a bright visible Herbig Ae/Be star, AB~Aurigae,
with a spectro-interferometer operating at optical wavelengths,
the VEGA spectrograph (\cite{vega}) installed at the CHARA Array (\cite{chara}).
This unique combination of spectral and spatially resolved information allowed us
to study the physical conditions for the gas producing the H$\alpha$ line on sub-AU scales.
Interpreting the results, however, required appropriate modeling of the emitted radiation.
Using a radiative transfer code, we obtained observational
insights on the generation mechanism of winds around AB~Aurigae, giving experimental support to the widely accepted, but not yet tested, theory of magneto-centrifugal acce\-leration.
In Sect.~2, we recall the present knowledge about the target.
In Sect.~3, we describe the observations and the data
processing. In Sect.~4, we present the radiative transfer code and
discuss the wind models adopted to interpret the observed H$\alpha$ visibilities and
spectrum.

\section{The target: AB~Aurigae}

AB~Aurigae (AB~Aur; A0Ve; d=144 pc) is the brightest
Herbig~Ae star in the northern hemisphere, and is often considered as
the prototype of the HAeBe class. Its photometric and spectral
properties   are   well  studied   over   a   wide  wavelength   range
(\cite{Bohm, Catala99, Gradya, VandenAncker}) and there is evidence of both accretion and
outflows  in AB  Aur (\cite{CatalaKunasz}, \cite{Garcia}, \cite{Wassel}).
The H$\alpha$ line exhibits a P-Cygni profile, which is a common
probe of stellar wind, and is known to be variable, especially in
the blue wing of the line. Among the alternatives evoked to explain the
observed variabi\-lity were a wind coming either from
the equatorial regions of the star or from  a circumstellar rotating
disk (\cite{Catala99}). The  wind in AB~Aur seems  to be
complex, probably controlled by a magnetic field, even if no magnetic detection of AB~Aur has been reported so
far (\cite{Alecian}). \\
AB~Aur is surrounded by a complex combination of gas-rich and
dust-dominated structures probed on large scales by imaging. Using
the HST, Grady et  al. (1999b) detected extended  nebulosity and
circumstellar material at all position angles  out to 1320  AU in the
R-band, whereas Fukagawa et al. (2004) detected a double
spiral structure at 200-450~AU in the H-band. Spatially resolved observations at high angular
resolution at near-infrared, thermal infrared, and millimetric wavelengths have shown the
presence  of  a  circumstellar  disk  (\cite{MG01}, \cite{Eisner04}),
rotating  at   non-Keplerian  velocities  (\cite{Pietu}). All these
observations agree on a low inclination for the system  (i.e.,
i$\leq$40$^\circ$), in disagreement with the value of 70$^\circ$ $\pm$ 20$^\circ$
deduced from spectroscopic studies (\cite{Catala99}). A strong  near-infrared  continuum emission  was
found inside the silicate dust sublimation radius (0.24~AU) and was
interpreted as due to a hot gaseous disk (\cite{Tannir08}). A localized
asymmetry  on spatial  scales  of 1-4  AU  has also  been detected  by
Millan-Gabet et al. (2006) who speculated that it  might result from
either loca\-lized viscous heating due to a gravitational instability in
the disk or from a close  stellar companion, in agreement
with   spectro-astrometric   observations   in   the   optical   lines
(\cite{Baines}).

\section{Observations and data processing}

Data were collected at the CHARA Array with the VEGA
 spectrometer recording spectrally dispersed fringes at
 visible wavelengths. Two pairs of telescopes were combined: W1W2 (whose projected baseline B$\sim$100~m) and S1S2 (B$\sim$34~m). Observations
 were performed around the H$\alpha$ line, between 610 and 700~nm,
 at the lowest spectral resolution of VEGA (R=1700). Observations of AB~Aur were sandwiched between those
 of a nearby calibration star (HD 29646) that was chosen to be bright enough
 ($m_{V}$  = 5.7) and to have  a  similar spectral  type (A2).  The
 observation log is given in Table~\ref{tab:log}.

\begin{table}[t]
\centering
\caption{Journal of AB~Aur observations on October 8 and 9, 2008.}
\label{tab:log}
\begin{tabular}{cccccc}
\hline Date & UT & Star & B (m) & PA ($^\circ$) & HA \\ \hline
2008-10-08 & 09:45 & HD 29646 & 33.9 & 1.66 & -1 h 27\\
2008-10-08 & 10:05 & AB Aur & 34.0 & 0.98 & -1 h 22\\
2008-10-08 & 10:24 & HD 29646 & 33.9 & -4.02 & -0 h 47\\
2008-10-08 & 10:45 & AB Aur & 34.0 & -4.94 & -0 h 40\\
\hline
2008-10-09 & 08:21 & HD 29646 & 86.1 & 120.9 & -2 h 32\\
2008-10-09 & 08:57 & AB Aur & 89.9 & 119.4 & -2 h 32\\
2008-10-09 & 09:25 & HD 29646 & 95.8 & 112.0 & -1 h 43\\
2008-10-09 & 09:49 & AB Aur & 99.0 & 110.3 & -1 h 25 \\ \hline
\end{tabular}
\end{table}

\begin{figure*}[t]
\centering
 \includegraphics[width=9.1cm, angle=0]{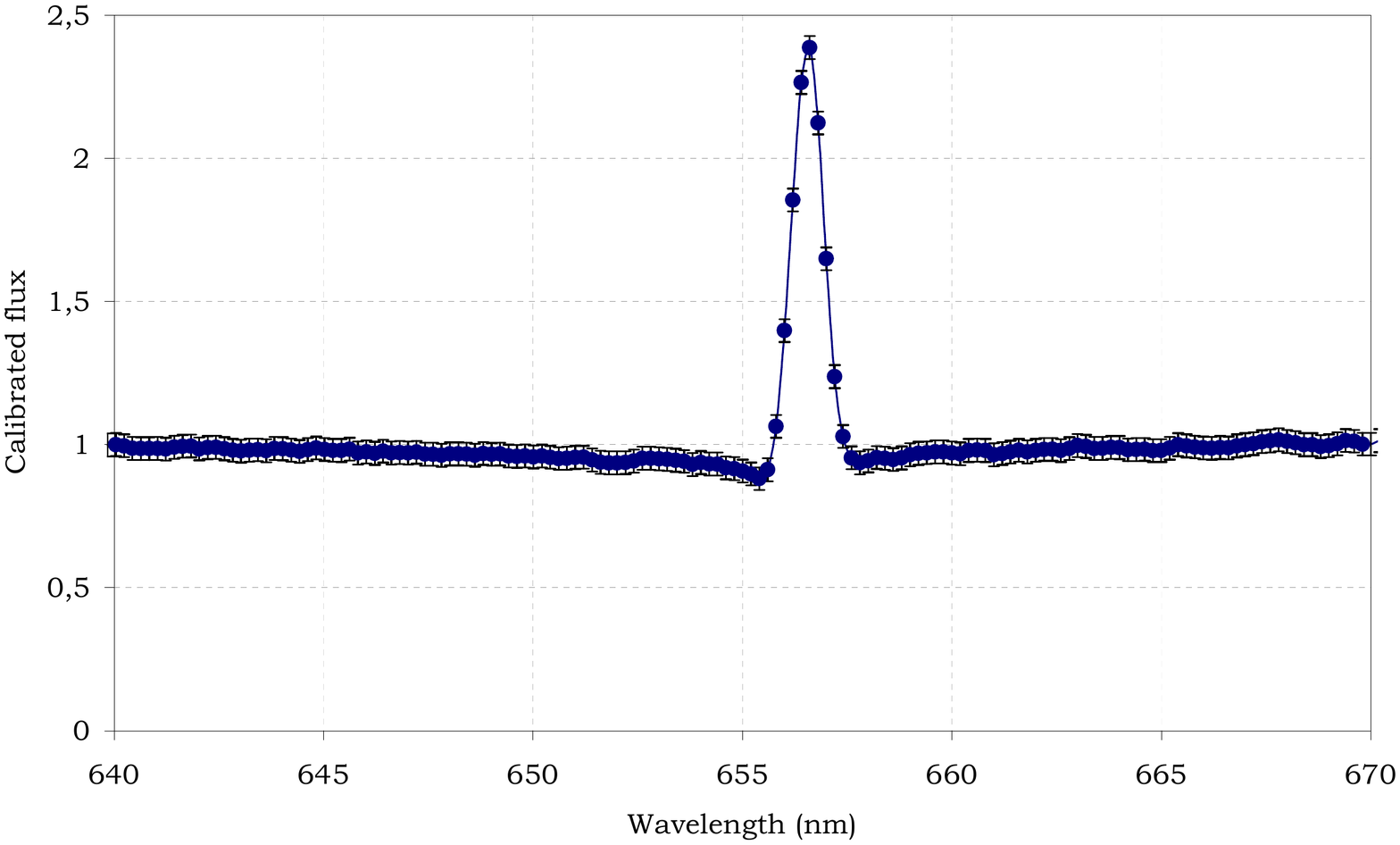}\includegraphics[width=9.1cm, angle=0]{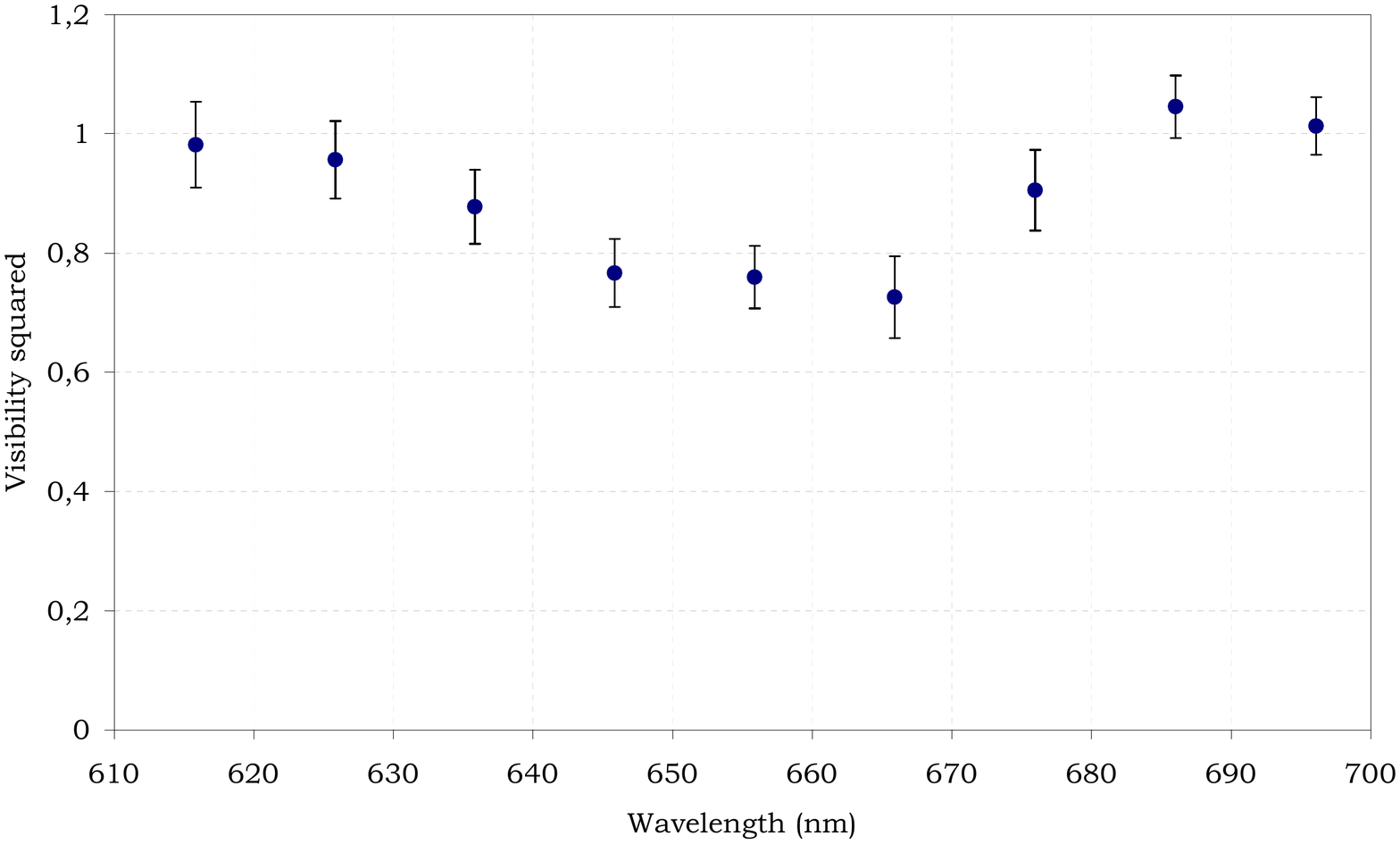}
\caption{\textbf{Left.} H$\alpha$ line recorded at R = 1700 with VEGA. \textbf{Right.} Calibrated squared visibility at R = 30 obtained with the S1S2 CHARA baseline.}
  \label{fig:data}
\end{figure*}

Each set of data was composed of observations following a sequence
calibrator-star-calibrator, with 10 files of 3000 short exposures per
observation. Each data set was processed
using the $C_{1}$ estimator and the VEGA data reduction pipeline
detailed in Mourard et al. (2009), with a large spectral band of 20.4~nm (shifted by steps
of 10.2~nm over the observing range) to reach a sufficient signal-to-noise ratio.
The resulting resolution of the spectral visibilities is R=30,
leading to an interferometric field of view of 240~mas. The poor signal-to-noise ratio meant that
the visibilities could not be processed for the
data recorded with W1W2. Only the squared visibilities correspon\-ding to S1S2 were computed in 9
spectral channels and calibrated using an angular diameter of 0.20$\pm$0.01~mas for the
calibrator HD~29646\footnotemark{}\footnotetext{$\textrm{http://www.jmmc.fr/searchcal\_page.htm}$} (Fig.~\ref{fig:data}-right).

The VEGA spectrum of AB~Aur exhibits an H$\alpha$ emission line with a clear P-Cygni
profile (Fig.~\ref{fig:data}-left). The line to continuum ratio is at
most about 2.5, the line FWHM is 6~\AA, and its emission peak is
slightly  red-shifted.  In addition,  the line  presents a  very broad
absorption from $\sim$640~nm until $\sim$665~nm, probably due to Stark
broadening in the absorption of the radiation by  nearly hydrostatic
cold    layers   located   just    above   the    star's   photosphere
(\cite{Felenbok83}). Due to the width of  the processing  spectral  window,
the squared visibilities at 646~nm, 656~nm, and 666~nm include all or part of the H$\alpha$ emission, and
the values of the squared visibility are between 0.72 and 0.75. Although they
are relative to continuum emission alone, the squared visibilities at 636~nm and 676~nm
are below unity, while those farther away from H$\alpha$ are consistent with 1.  AB~Aur is
thus clearly resolved by the S1S2 baseline both in the H$\alpha$
line and in a part of the visible conti\-nuum.

\section{Modeling and discussion}

\begin{figure*}[t]
\centering
 \includegraphics[width=9.1cm, angle=0]{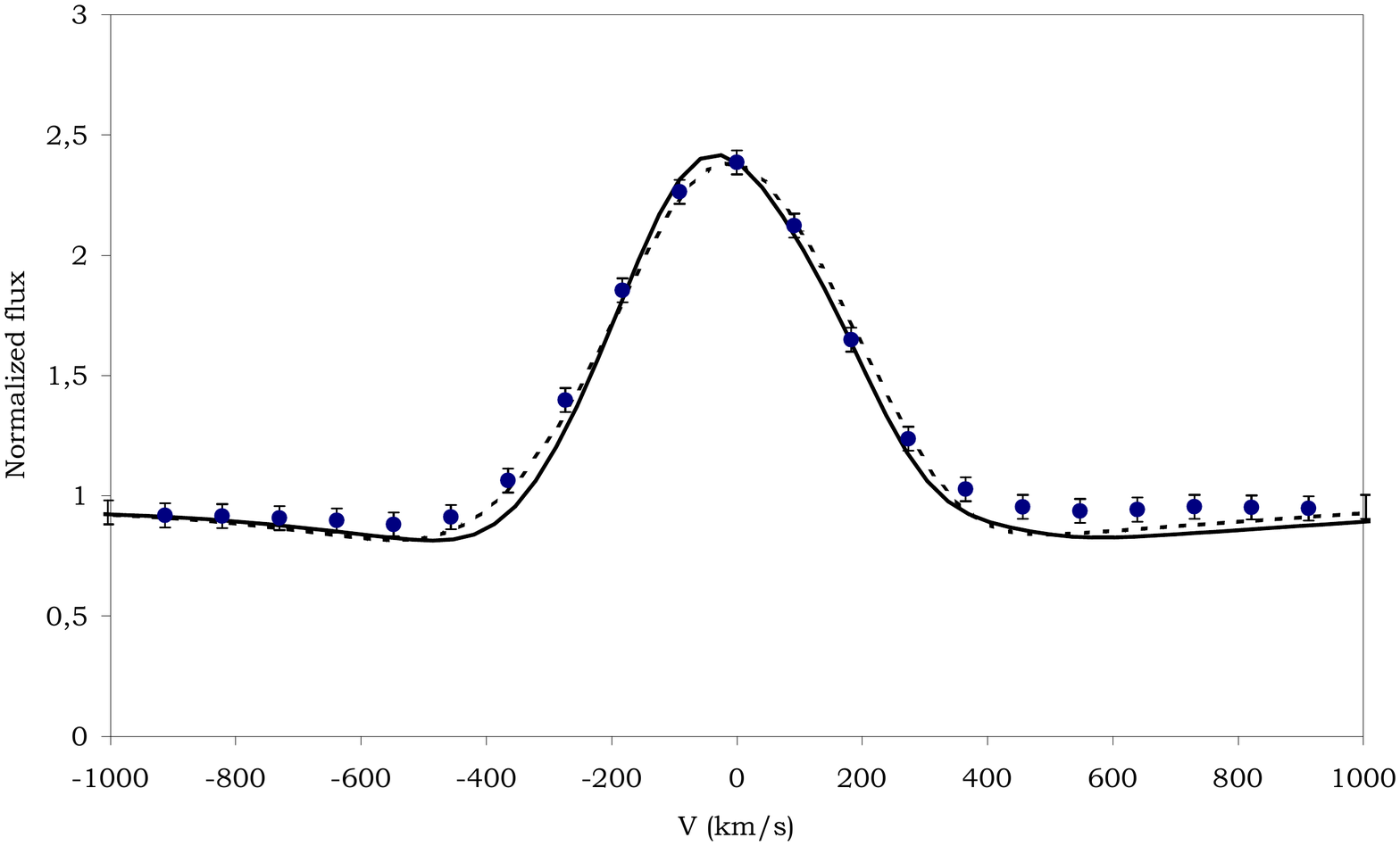}
 \includegraphics[width=9.1cm, angle=0]{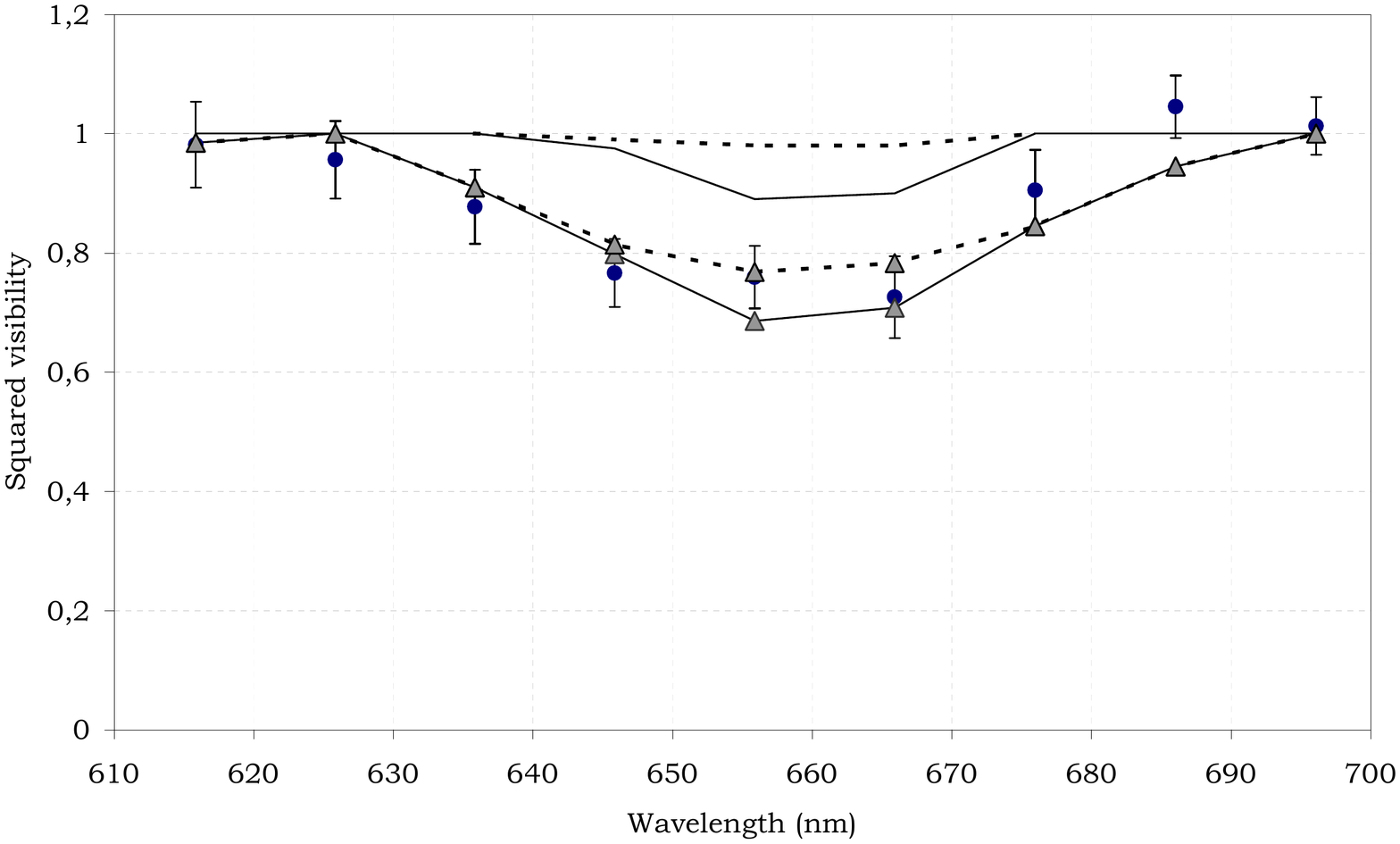}
  \caption{Best model predictions compared to the VEGA spectrum (left)
 and spectral  visibilities (right), both  in full circles  with error
 bars. Dashed lines  correspond to the X-wind model  and full lines to
 the disk-wind  one. For the squared visibilities,  triangles denote the
 model prediction including both the wind and the contribution of the
 asymmetric reflection nebula.}
  \label{fig:model-disc-binary}
\end{figure*}

The spectroscopic P-Cygni feature in AB~Aur shows that the line
is formed at least in part in a wind. Therefore, we focused on a wind
model to constrain the properties of the emitting region. To simultaneously
reproduce both the line profile and the spectral visibilities, we have
adapted two numerical codes in spherical symmetry for the production and transport
of hydrogen (H) lines (\cite{Rajabi}). The first one, RAMIDUS, calculates the level populations of H in a moving fluid
for assigned radial dependencies of the basic wind parameters (density, velocity, tempe\-rature).
The code uses the Sobolev approximation and the escape probability method (\cite{Castor70}). In cascade, the PROFILER  code calculates synthetic images, line profiles and visibilities of H lines
 for various geometrical configurations. By selecting limited portions of
 the spherical winds, we approximated 2-D axisymmetric models. We did not
 include accretion in the models.

\begin{figure}[t]
\centering
\includegraphics[width=0.35\textwidth]{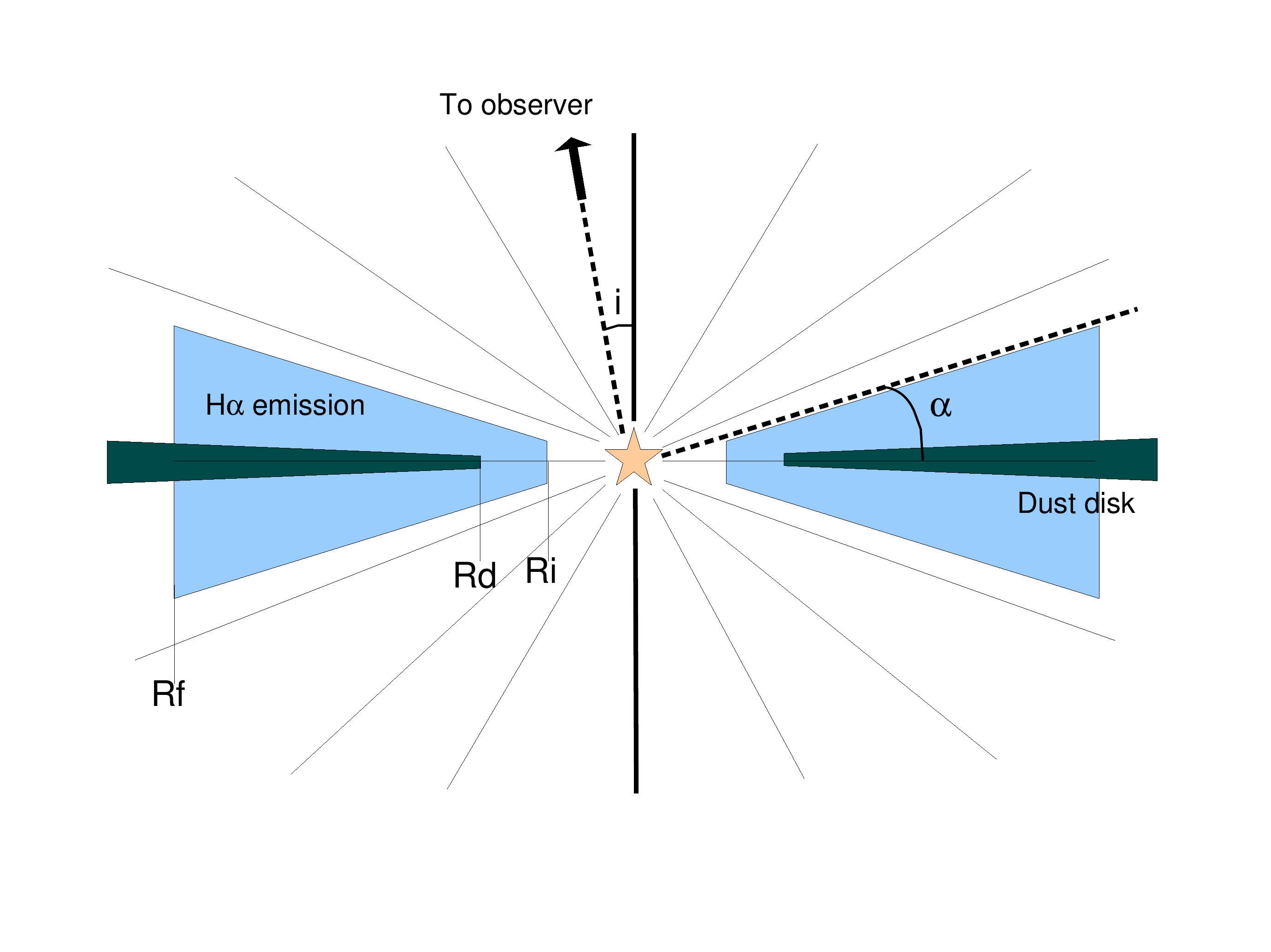}
  \caption{Geometry adopted for the empirical X-wind and disk-wind
mo\-dels described in Table~\ref{tab:model}.  R$_i$, R$_d$, and R$_f$ are
the corotation radius, dust sublimation radius, and extension of the wind,
respectively.}
  \label{fig:model}
\end{figure}

A full spherical wind providing matter along the line of sight can be
readily  excluded,  because it  would  produce  too much  blue-shifted
absorption  with respect to the observed  P-Cygni  feature that  is
shallow but not very deep (normalized flux$\geq$0.88). Even
 if the issue of the inclination angle of the AB Aur system is still
alive (see Sect. 2), we chose a wind geometry in agreement with a small
inclination (i$\leq40^{\circ}$), and thus excluded a double cone
geometry, reminiscent  of a bipolar  jet, which would lead, for small inclinations,
to a line profile with two distinct emission features instead
of the observed single peak. Inspired by the magneto-centrifugal (MC)
scenario for  the acceleration of  stellar jets, we assume
that, on the observed scales (much less than 2-3 AU from the star/disk),
the wind occupies a region delimited by a  flattened torus defined by
the pale gray region in Fig.~\ref{fig:model}.  This configuration can
simulate       the      base       of       either      a       X-wind
(\cite{Camenzind90}, \cite{Shu95}) or a disk-wind (hereafter D-wind;
\cite{Ferreira97},   \cite{KoniglPudritz})   accelerated   along   the
magnetic field  lines anchored  in the inner  few AU of  the accretion
disk.  In this region the wind has a wide angle aperture, as it has
not yet reached the  Alfv\`en  surface, after  which  magnetic
collimation occurs and the wind collimates into a jet.
In  Fig.~\ref{fig:model}, R$_i$=0.02~AU  is the  corotation radius,
R$_d$=0.24~AU is the inner radius of the optically thick dusty disk,
R$_f$=2.3~AU is  the extension  of the  wind, and
$\alpha $ is the angle between the disk mid-plane and the
surface of the wind, constrained to be  $\leq 60^{\circ}$ for MC acceleration
to occur (\cite{Spruit96}).  In our X-wind model, all  the material in
the flow comes  from the region located very close to the corotation radius
and has  a constant  mass outflow rate  $\dot{\rm{M}}_w$ through  the torus
section, while for the D-wind $\dot{\rm{M}}_w$ progressively increases from R$_i$ to
about 1~AU, to simulate that the wind originates in an extended
region of the disk. The wind velocity is assumed to be $V(r)=V_{0}+V_{f}(1-(R_{\star}/r)^2)$,
where $V_{0} =$ 20 km.s$^{-1}$ is the initial velocity, $r$ the
radial distance from the star, $R_{\star}$ the stellar radius, and
$V_{f}$ the terminal velocity.  The density profile is dictated by $V(r)$ and $\dot{\rm{M}}_w (r)$
through the mass continuity equation, while the gas temperature profile
T$(r)$, which reaches a maximum of 12~000 K, is chosen in accordance with the theory of wind heating illustrated by
Shang et al. (1998) and Garcia et al. (2001) for the X-wind and D-wind, respectively.
Modeling of the Stark absorption in the extended line wings has been
included using a template spectrum from Kurucz~(1979). 

 Regarding the modeling of the line profile, both an X-wind and
 a D-wind could provide a good fit (Fig.~\ref{fig:model-disc-binary}-left)
 for the model parameters collected in Table~\ref{tab:model}. To reproduce
 at the same time the line width and the absence of a strong
 blue-shifted absorption, the best predictions for the X-wind
 have nearly a null inclination ($i$) and a large opening angle ($\alpha$)
 of 60$^{\circ}$. For the D-wind model, since the disk hides part of the high
  velocity material in the red-shifted wing, the best predictions are
   obtained with a small inclination (20$^{\circ}$) and a smaller
   opening  angle of 35$^{\circ}$ to avoid  too  much  blue-shifted
   absorption. Both models lead to a small inclination in agreement with the
   previous interferometric determinations. As pointed out by  Pogodin (1992), a change in the
   opening angle of the wind ($\alpha$) affects the depth of the
   blue-shifted absorption. Concerning the mass loss rate $\dot{\rm{M}}_w$, our best prediction
   is of the same order as previous determinations from line profile analysis
   (\cite{CatalaKunasz}) for the X-wind but is significantly higher for the D-wind.
   In fact, our various simulations clearly show that the estimate of
   $\dot{\rm{M}}_{\rm  w}$ is very dependent on the velocity law and the
   temperature profile considered in the wind model, and on the scale on which the wind
   is accelerated. Since the geometry of our D-wind model strongly deviates from a spherical wind, the value of
   $\dot{\rm{M}}_{\rm  w}$ obtained for this model cannot
   be directly compared to the previous determinations of $\dot{\rm{M}}_w$.

\begin{table}[t]
\centering
\caption{Parameters of  the models, with N$_{\rm{H},ej}$ the average
  hydrogen density at the wind footpoints. } 
\label{tab:model}
\begin{tabular}{c c c c c c c}
\hline
Model  &   $i$   &  $\alpha$   &  V$_{\rm{f}}$
[km/s] & $\dot{\rm{M}}_w$ [M$_{\odot}$/yr]& N$_{\rm{H},ej}$ [cm$^{-3}$]\\
\hline
  X-wind & 0$^\circ$ & 60$^\circ$  & 410 &  3.7 10$^{-8}$ & 4.3~$10^{10}$ \\
  D-wind & 20$^\circ$ & 35$^\circ$ & 450 &  2.2  10$^{-7}$ & 7.7~$10^{8}$\\
  \hline
\end{tabular}
\end{table}

Regarding the modeling of V$^2$, the X-wind  provides only a negligible decrease in H$\alpha$
($V^2$ = 0.98) with           respect            to           the           continuum
(Fig.~\ref{fig:model-disc-binary}-right). In this  case, the
emission is concentrated  at R$_i$, while it is  much more extended in
the D-wind model that produces a larger decrease in $V^2$ (down to 0.89).
In both cases, however, the observed visibility variation could not be
modeled  with the  wind alone so we considered an additional component
producing an extended continuum emission. This component
could come from the extended reflection nebulosity around the system,
already reported by several authors (\cite{Gradyb, Fukagawa, MG06}).
In practice, we used this secondary source to reproduce the modulation
of the visibility in the continuum. Our wind models combined with this additional emission provided
a good fit of the visibilities. For instance, Fig.~\ref{fig:model-disc-binary}-right
corresponds to a secondary source separated from the star by 38.2 mas
(i.e. 5.5 AU) contributing to 7\% of the total flux. Because of the large
number of free parameters and the fact that we only
have measurements along one baseline direction, we found many
different configurations for the secondary emission that could fit, for
both  wind  models, the  observed visibilities (e.g. a much larger separation of
591.2 mas).

In conclusion, thanks to the unique capabilities of VEGA/CHARA,
we managed for the first time to record simultaneously a spectrum at a resolution of 1700 and a visibi\-lity spectrum in the visible range on a target as faint as $m_{V}$ = 7.1. Despite the limited signal-to-noise ratio of our data, they allowed us to rule out a spherical wind and to make realistic predictions on the nature of the H$\alpha$ emission region at sub-AU, and our data are compatible with a magneto-centrifugal mechanism for the production of the wind. It was difficult, however, to determine the exact morphology of the wind and disentangle the X-wind and the D-wind because of the extended nebu\-losity. Moreover, all models neglect the contribution of mass accretion to the H-line emission, so self-consistent models of winds including this process are needed. Complementary observations with larger CHARA baselines are required to place more stringent constraints on the complex environment and the photosphere contribution of AB~Aur that can be resolved in the visi\-ble with VEGA. We also expect to take advantage of the new instrumental facilities of CHARA to study other bright HAeBe and provide promising insights on the accretion/ejection mechanisms in intermediate-mass young stars.

\begin{acknowledgements}
VEGA is a collaboration between CHARA and OCA-LAOG-CRAL-LESIA supported by the French programs PNPS, ASHRA, by INSU, by the R\'egion PACA, and by the OCA and CHARA technical teams. The CHARA Array, operated by Georgia State University, was built with funding provided by the National Science Foundation (NSF), Georgia State University (GSU), the W. M. Keck Foundation, and the David and Lucile Packard Foundation. Array operations are supported by NSF (grant AST0606958) and by the office of the Dean of the College of Arts and Science at GSU. M.B. acknowledges funding from INAF (grant ASI-INAF  I/016/07/0). S.R. acknowledges funding from European Community FP6 (MCRTN 005592) and FP7 (GA 226604). We warmly thank the referee whose comments helped us clarify some important issues. We thank C. Giovanardi and P. Garcia for fruitful discussions. This research  has made  use of  the SearchCal service of the  Jean-Marie Mariotti Center and  of CDS Astronomical Databases SIMBAD and VIZIER.
\end{acknowledgements}

{}

\end{document}